\documentclass[12pt]{article}
\usepackage{tikz}
\usetikzlibrary{arrows,decorations.markings,mindmap,patterns}

\newlength{\oldoddsidemargin}
\oldoddsidemargin=\oddsidemargin \oddsidemargin=\evensidemargin
\evensidemargin=\oldoddsidemargin

\makeatletter
\def\cleardoublepage{\clearpage\if@twoside \ifodd\c@page\else
   \hbox{}
   \thispagestyle{empty}
   \newpage
   \if@twocolumn\hbox{}\newpage\fi\fi\fi}
\makeatother \clearpage{\pagestyle{plain}\cleardoublepage} 
\usepackage{tabularx}
\usepackage{longtable}
\usepackage{multirow}
\usepackage{graphicx, color}
\usepackage{subfigure}
\usepackage{amsmath}
\usepackage{amsfonts}
\usepackage{amssymb}
\usepackage[english,german]{babel}
\usepackage{bbding}

\usepackage{float}
\usepackage{dsfont}   % unit matrix symbol (\mathds{1}), or real numbers \mathds{R} ... 
\usepackage{hyperref}
\usepackage{mathrsfs} 
\usepackage{makeidx}
\usepackage{listings}
\makeindex
\usepackage{setspace}
\newcolumntype{C}[1]{>{\centering\arraybackslash}b{#1}}
\newcolumntype{D}[1]{>{\centering\arraybackslash}b{#1}}
\newcolumntype{R}[1]{>{\raggedleft\arraybackslash}b{#1}}                                                                                                                                                                                                                                                                                                                                                                                                                                                                                                                                                                                                                                                                                                                                                                                                                                                                                                                                                                                                                                                                                                                                                                                                                                                                                                                                                                                                                                                                                                                                                                                                                                                                                                                                                                                                                                                                                                                                                                                                                                                                                                                                                                                                                                                                                                                                                                                                                                                                                                                                                                                                                                                                                                                                                                                                                                                                                                                                                                                                                                                                                                                                                                                                                                                                                                                                                                                                                                                                                                                                                                                    
\newcolumntype{M}[1]{>{\centering\arraybackslash}m{#1}}

\def\d{\mathrm d}

\def\bea{\begin{eqnarray}}
\def\eea{\end{eqnarray}}

\def\I{\mathrm{i}}

\def\bsm{\left( \!\begin{smallmatrix}}
\def\esm{\end{smallmatrix} \!\right)}

\renewcommand{\Re}{\mathrm{Re}}
\renewcommand{\Im}{\mathrm{Im}}
\newcommand{\mac}[1]{\mathcal{#1}}

\newcommand{\nn}{\nonumber}

\numberwithin{equation}{section}

\renewcommand{\a}{\alpha}

\newcommand{\Th}{\Theta}

\newcommand{\pt}{\partial}

\newcommand{\be}{\begin{equation}}
\newcommand{\ee}{\end{equation}}
\newcommand{\ba}{\begin{align}}
\newcommand{\ea}{\end{align}}

\def\cR{\mac{R}}
\def\cD{\mac{D}}

\def\bcD{\bar{\mac{D}}}
\def\Phid{\Phi^\dagger}

\begin{document}
\allowdisplaybreaks
\selectlanguage{english}

\begin{titlepage}

\title{DBI Inflation in $\mac{N}=1$ Supergravity {\LARGE \\[.5cm]  }}
\author{{Michael Koehn${}^{1}$, Jean-Luc Lehners${}^{1}$ and Burt A. Ovrut${}^{2}$}\\[5mm]
{\it ${}^{1}$ Max-Planck-Institute for Gravitational Physics} \\
{\it Albert Einstein Institute, 14476 Golm, Germany}\\[4mm]
{\it  ${}^{2}$ Department of Physics, University of Pennsylvania} \\
   {\it Philadelphia, PA 19104--6396}}
\date{}
\maketitle

\begin{abstract} 

\let\thefootnote\relax\footnotetext{michael.koehn@aei.mpg.de,~~~jlehners@aei.mpg.de,~~~ovrut@elcapitan.hep.upenn.edu} 
It was recently demonstrated that, when coupled to ${\cal{N}}=1$ supergravity, the Dirac-Born-Infeld (DBI) action constructed from a single chiral superfield has the property that when the higher-derivative terms become important, the potential becomes negative. Thus, DBI inflation cannot occur in its most interesting, relativistic regime. In this paper, it is shown how to overcome this problem by coupling the model to one or more additional chiral supermultiplets. In this way, one obtains effective single real scalar field DBI models with arbitrary positive potentials, as well as multiple real scalar field DBI inflation models with hybrid potentials.
\vspace{.3in}
\noindent
\end{abstract}

\thispagestyle{empty}

\end{titlepage}

\tableofcontents

\section{Introduction}

Inflation is a possible solution to the flatness and horizon puzzles of standard big bang cosmology. It was discovered more than 30 years ago \cite{Guth:1980zm,Linde:1981mu,Albrecht:1982wi}, and is considered by many as the leading cosmological model of the early universe. This is due in large part to its ability to generate nearly scale-invariant density perturbations at the same time as addressing the above-mentioned puzzles. However, inflation is not unique in this regard. For example, during an ekpyrotic phase \cite{Khoury:2001wf,Lehners:2007ac,Lehners:2008vx}, where the universe contracts very slowly, the same big bang puzzles can be addressed and nearly scale-invariant density perturbations can be generated\footnote{In these models, one must also understand the transition from a contracting to an expanding phase--this remains an open issue, but see \cite{Turok:2004gb,Buchbinder:2007ad,Buchbinder:2007tw,Lehners:2011kr}.}. It follows, therefore, 
that to understand the actual history of our universe, we must make progress in two directions. On one hand, it is important to work out the detailed predictions of the various models of the early universe--in particular, the 
different predictions they make regarding the non-Gaussian features in the primordial density perturbations \cite{Maldacena:2002vr,Chen:2010xka,Buchbinder:2007at,Koyama:2007if,Lehners:2007wc,Lehners:2008my,Lehners:2009ja,Lehners:2009qu}. On the other hand, it is imperative to develop the microphysical structure of the various cosmological models. In this paper, we will be mainly interested in this second aspect--with the focus on inflationary models. 

Our aim is to study inflationary theories with higher-derivative kinetic actions in the context of four-dimensional, ${\cal{N}}=1$ supergravity\footnote{A study of DBI inflation in global supersymmetry (with an added Einstein-Hilbert term) was performed in \cite{Sasaki:2012ka}.}. Although our work will be purely within this supergravity context, the motivation stems from string theory. There, the dynamics of D-branes and M5-branes are described by the Dirac-Born-Infeld (DBI) action \cite{Leigh:1989jq}\footnote{The effective description in terms of the DBI action is valid at arbitrary velocity, but only as long as the proper acceleration of the branes is small.}. This action is unusual in that it contains higher-derivative terms which are essential to understanding its dynamics\footnote{Higher-derivative terms involving the extrinsic and intrinsic brane curvatures--such as those discussed in \cite{Khoury:2012dn,Ovrut:2012wn}--can arise as well. We will not consider these couplings here, but note that they might be significant in certain applications.}. Furthermore, interactions between branes (and anti-branes) can generate an effective potential \cite{Dvali:1998pa,Lima:2001nh,Buchbinder:2002pr,Kachru:2003sx,Gray:2007zza}. In such a setting, inflationary models based on the DBI action, in which the inflaton field is identified with a position modulus of the brane, have been constructed and shown to lead to interesting observational predictions--such as equilateral non-Gaussianities \cite{Silverstein:2003hf,Alishahiha:2004eh}. These models have mainly been analyzed in non-supersymmetric effective field theory. However, realistic string compactifications typically preserve minimal supersymmetry in four dimensions--see, for example \cite{Braun:2005nv,Lukas:1998yy}. It is of interest, therefore, to re-formulate these models within the context of four-dimensional, $\mac{N}=1$ supergravity. 

In a recent paper \cite{Koehn:2012ar}, we developed a formalism for coupling chiral supermultiplets with higher-derivative kinetic terms to supergravity. Restricting to a single chiral superfield, we constructed a supergravitational generalization of the single real scalar DBI action. This supergravity  theory then contains the DBI action of two real scalar fields--the constituents of the lowest component of the chiral supermultiplet--along with a specific potential energy. In the process, however, we discovered that when the higher-derivative terms become significant, the potential energy necessarily becomes negative--regardless of the form of the superpotential. Thus, with a single chiral supermultiplet, DBI inflation cannot occur! In this paper, we will show how this restriction can be overcome by coupling the supergravity DBI theory to one or more additional chiral superfields--each, however, with canonical two-derivative kinetic terms. Such couplings can lead to positive, inflationary potentials via the elimination of the new auxiliary fields. The required couplings are similar, and in some cases identical, to those previously considered in several two-derivative inflationary models in supergravity \cite{Kawasaki:2000yn,Kallosh:2010xz,Lazarides:1995vr}. However, in the higher-derivative context, they lead to a number of new features, and to different predictions for cosmological observations.  

We have two main results. 1) Within the context of ${\cal{N}}=1$ supergravity, we provide a method for obtaining DBI inflation for a single real scalar component of a chiral superfield with an arbitrary potential energy. This is accomplished 
both when the higher-derivative terms are negligible and, more importantly, in the relativistic regime where the higher-derivative terms are dominant. We achieve this by coupling the single chiral superfield DBI theory to one additional chiral supermultiplet--with two-derivative kinetic energy, constrained K\"ahler potential and specific holomorphic couplings. 2) We show how one can obtain multi-real-field DBI models with positive potentials. There are two possibilities here. First, within the context of the models just discussed one can allow the scalar superpartner of the inflaton field to fully participate in the dynamics. This is accomplished by easing restrictions on the K\"ahler potential. In this case, the potential for the second real scalar field is automatically determined. Second, and more generally, one can couple the supergravity DBI theory to two or more additional chiral supermultiplets--in which case there is more freedom in constructing multi-field potentials. The multi-real-scalar-field models are of clear phenomenological interest, since they can each be compatible with current observational data while making predictions that are testable in the near future \cite{Langlois:2008qf,Kidani:2012jp}.
 
The plan of the paper is the following. In Section \ref{SectionReview}, we review the construction of single chiral superfield DBI actions in ${\cal{N}}=1$ supergravity. This reveals that, in the relativistic regime, the potential for both real component scalars in the  DBI action is negative and thus prohibits inflation from occurring. In Section \ref{SectionSingle}, we show how the inclusion of a second chiral supermultiplet modifies this conclusion. In fact, via a judicious choice of both the 
K\"ahler potential and superpotential, this allows arbitrary positive potentials to be constructed for a single real DBI scalar field--while simultaneously fixing the remaining three real scalars. In the beginning of the next section, we briefly discuss how this theory can be modified so that both real component scalars of the DBI superfield become dynamical. In Section \ref{SectionMulti}, we introduce a third chiral superfield. This allows us to construct a more general class of multi-field models of DBI inflation in supergravity, including, for example, models with inflationary potentials of the hybrid type. We conclude in Section \ref{SectionConclusions}.

\section{Higher-Derivative Kinetic Terms in Supergravity} \label{SectionReview}

In \cite{Koehn:2012ar}, we showed how to couple chiral superfields with higher-derivative kinetic terms to four-dimensional ${\cal{N}}=1$ supergravity\footnote{Also see \cite{Farakos:2012qu}, where related results were obtained. Earlier work of interest includes \cite{Khoury:2010gb,Khoury:2011da,Buchbinder:1988yu,Buchbinder:1994iw,Banin:2006db,Brandt:1993vd,Brandt:1996au,Antoniadis:2007xc}.}. Since we are interested in cosmological applications, fermionic component fields will be ignored throughout. The construction takes place in curved superspace, which is the most natural setting for writing actions invariant under local supersymmetry transformations. A chiral superfield $\Phi$ then admits the expansion 
\be
\Phi = A  + \Th^\a \Th_\a F,
\ee 
where $A$ is a complex scalar field and $F$ is a complex auxiliary field. The $\Th$ coordinates are Grassmann-valued and carry local Lorentz indices ($\a$ denotes the index of a two-component  Weyl spinor). They extend ordinary spacetime to curved superspace, and are defined precisely so that $A$ and $F$ arise as the components of $\Phi$ in the above expansion. In curved superspace, supersymmetric Lagrangians can be constructed from the chiral integrals
\be
\int \d^2 \Th (\bcD^2 - 8 R) L,
\ee
where $L$ is a scalar, hermitian function. The chiral projector in curved superspace is $\bcD^2 - 8R,$ where $\bcD_{\dot\a}$ is a spinorial component of the curved superspace covariant derivative $\cD_A=\{ \cD_a,\cD_\a,\bcD_{\dot\a}\}$ and $R$ is the curvature superfield.  In its component expansion, $R$ contains the Ricci scalar $\cR$ as well as the auxiliary fields of 
supergravity--namely a complex scalar $M$ and a real vector $b_m.$ The purely bosonic components in the $\Th$ expansion of $R$ are
\be
R = -\frac16 M + \Th^2 \big( \frac{1}{12}\cR -\frac19 MM^* - \frac{1}{18} b_m b^m + \frac16 \I {e_a}^m \cD_m b^a\big) \ .
\ee
Another superfield that we will need is the chiral density $\mac{E}$ with expansion
\be
2 \mac{E} = e (1 - \Th^2 M^*),
\ee
where $e$ is the determinant of the vierbein. Note that the tangent space Lorentz indices $A=\{ a,\a,\dot\a\}$ are related to the spacetime indices $M=\{ m,\mu,\dot\mu\}$ via the supervielbein ${E_M}^A$ and its inverse, with ${E_m}^a ={e_m}^a$ being the ordinary vierbein. For a complete discussion of curved superspace we refer the reader to \cite{Wess:1992cp}.
 
The supergravity theory of chiral supermultiplets with higher-derivative kinetic terms is defined via the Lagrangian
\bea
\mac{L} &= &\int \d^2\Theta 2\mac{E}\Big[ \frac{3}{8}(\bcD^2-8R) e^{-K(\Phi^i,\Phi^{\dagger k*})/3}+W(\Phi^i)\Big]+h.c. \nn \\
&& -\frac18 \int \d^2\Theta 2\mac{E} (\bcD^2-8R) \cD \Phi^i \cD \Phi^j \bcD \Phi^{\dagger k*} \bcD \Phi^{\dagger l*} \, T_{ijk*l*}+h.c. \label{Action}
\eea
The first two terms contain the  K\"{a}hler potential $K(\Phi^i,\Phi^{\dagger k*}),$ which is an hermitian  function of the chiral 
superfields $\Phi^i$ (where index $i=1,2,\dots$ enumerates the fields) and the superpotential, given by the holomorphic function $W(\Phi^i)$. By themselves, these terms lead to ``normal'' two-derivative kinetic energy and a potential for the scalar superfields coupled to canonical supergravity. The final term, however, describes chiral superfields with higher-derivative kinetic energy. $T_{ijk*l*}$ is a tensor superfield that is hermitian and symmetric in the indices $i,j$ as well as in $k^*,l^*$. It contains an arbitrary real function of the chiral superfields and their covariant space-time derivatives ${\cal{D}}_{m}$, with all such indices contracted. 
Here, we will be interested in the case where only one of the chiral superfields, namely $\Phi^1 \equiv \Phi,$ has a higher-derivative action--the generalization to many superfields with higher-derivative actions being straightforward. In that case, $T_{ijk*l*}$ effectively reduces to a single arbitrary function $T$ of $\Phi,\Phid$ and their spacetime derivatives. 

By choosing this function appropriately, one can write a supergravity version of the single real scalar field DBI action. It turns out that we need to consider a K\"{a}hler potential with the property
\be
\frac{\pt^2 K}{\pt \Phi \pt \Phid}| = K_{,AA^*} = 1
\ee and a tensor superfield \cite{Rocek:1997hi,Koehn:2012ar}
\be
16T =  \frac{f(\Phi,\Phid) }{1 + f \pt \Phi \cdot \pt \Phid e^{K/3} + \sqrt{(1 + f \pt \Phi \cdot \pt \Phid e^{K/3})^2 - f^2  (\pt \Phi)^2 (\pt \Phid)^2 e^{2K /3}}} \label{TDBI}.
\ee
Here $f(\Phi,\Phid)$ is an arbitrary hermitian function and we have used the notation that $\pt \Phi \cdot \pt \Phid = g^{mn} \cD_m \Phi \cD_n \Phid$. In a brane setting, the lowest component of the $f$ function can be identified with the warp factor of the direction in which the brane moves.  Performing the $\d^2\Th$ integral in the Lagrangian (\ref{Action}) picks out the $\Th^2$ component of the integrand. A feature of chiral supergravity is that, after performing this integration, one does not end up in Einstein frame. Rather, one has to perform a Weyl rescaling of the fields first, with the vierbein transforming as
\be
{e_m}^a \rightarrow {e_m}^a e^{K/6}.
\ee
Note that this rescaling also removes the factors of $e^{K/3}$ in (\ref{TDBI}). Then, after eliminating the auxiliary fields $b_m,M$ of the supergravity multiplet, the Lagrangian reduces to
\bea
\frac1e {\cal L} &=& -\frac12 \cR + 3 e^{K} |W|^2  \nn \\ && -\frac{1}{f} \left( \sqrt{1 + 2 f \, \pt A \cdot \pt A^* + f^2 \, (\pt A \cdot \pt A^*)^2 - f^2 \, (\pt A)^2 (\pt A^*)^2} -1 \right)  \nn \\ 
&& +  e^{K/3} |F|^2 + \frac{}{}e^{2K/3}\big(F (D_A W) + F^{*} (D_A W)^*\big)  \label{ActionComponents1}   \\ &&  -32 \frac{}{}e^{K/3} |F|^2 \pt A \cdot \pt A^* \, {\cal{T}}  + 16 e^{2K/3} |F|^4 \, {\cal{T}} \ . \nn
\eea
Here $\mac{T},$ which is the Weyl rescaled lowest component of $T,$ is given by
\be
16 \mac{T} = \frac{f}{1 + f\, \pt A \cdot \pt A^*  + \sqrt{(1 + f\, \pt A \cdot \pt A^*)^2 - f^2 \, (\pt A)^2 (\pt A^*)^2}}
\ee
with $f=f(A,A^*)$. 
The second line of (\ref{ActionComponents1}) can be recognized as the DBI action for the {\it two} real scalar fields $\phi,\xi$ that make up the complex scalar $A$ \cite{Rocek:1997hi}. That is, the simplest ${\cal{N}}=1$ supergravity generalization of the single real scalar DBI action naturally produces a DBI theory for both real scalar component fields. As can be seen from the action, when the fields depend only on time there exists an upper bound on the velocity of $A$ given by
\be
|\dot{A}|^2 \leq \frac{1}{2f}. \label{SpeedLimit}
\ee
The so-called {\it relativistic regime} corresponds to the situation where this bound is (almost) saturated. Models of DBI inflation \cite{Silverstein:2003hf} exploit this inequality. As the brane moves towards a region of large $f,$ the scalars are automatically constrained to move slowly--allowing for inflation to occur on potentials that would otherwise be too steep.

In the above Lagrangian, the auxiliary field $F$ has not yet been eliminated. Its equation of motion is algebraic, and given by
\be
F+ e^{K/3} (D_A W)^* + 32 F \, \mac{T} (e^{K/3} |F|^2 - \pt A \cdot \pt A^*) = 0. \label{EqofMotionFDBI1} 
\ee
Interestingly, this is a cubic equation. Thus, $F$ admits up to three solutions. In our previous paper \cite{Koehn:2012ar}, we showed that one of these solutions, which we termed the {\it ordinary branch}, is directly related to the usual solution for $F$ that one obtains in the absence of higher-derivative terms. In this paper, we only consider this branch. The remaining two branches lead to entirely different theories, which are not continuously connected to the ordinary branch as the higher-derivative terms become small. 
The ordinary branch solution for $F$ is given by
\be
F = F_+ + F_-,
\ee 
where
\bea
&& F_\pm = \left( -\frac{q}{2} \pm \left( (\frac{q}{2})^2 + (\frac{p}{3})^3 \right)^{1/2} \right)^{1/3}, \\
&& q =  \frac{1}{32 \mac{T}} \frac{(D_A W)^{*2}}{D_A W}, \qquad p = e^{-K/3} \frac{(D_A W)^*}{D_A W} \left( \frac{1}{32 \mac{T}} -\pt A \cdot \pt A^* \right). \nn
\eea
When $f$ is small, so is $\mac{T}$ and $F$ approaches the usual solution
\be
F \approx -e^{K/3} (D_A W)^* . \qquad (f \, \rm{small})
\ee
In this {\it  non-relativistic} limit, after substituting for $F$ one obtains the usual potential
\be
V_{\rm{non-rel.}} = e^K \big(\, |D_A W|^2  - 3 |W|^2\big). \label{PotentialNonRelativistic1}
\ee
Note that this expression is only valid as long as the higher-derivative terms in $A$ are irrelevant.
 
More interesting for our purposes is the {\it relativistic limit}, where $f$ is large and $|\dot{A}|^2$ correspondingly small, with $\mac{T}\approx f/8$. In that case, the solution for $F$ approaches
\be
F \approx  - \left( \frac{(D_A W)^{*2}}{4f \, D_A W} \right)^{1/3} . \qquad (f \, \rm{large})
\ee
After substituting for $F$ in the relativistic limit, the Lagrangian becomes
\bea
\frac1e {\cal L}_{\rm{rel.}} &=& -\frac12 \cR + 3 e^{K} |W|^2 -\frac{3}{2} \frac{e^{K} |D_A W|^2}{\big(4f\, e^{K} |D_A W|^2\big)^{1/3}} \label{ActionRelativistic1} \\ && -\frac{1}{f} \left( \sqrt{1 + 2 f \, \pt A \cdot \pt A^* + f^2 \, (\pt A \cdot \pt A^*)^2 - f^2 \, (\pt A)^2 (\pt A^*)^2} -1 \right)    \nn \\
&&+{\cal {O}}(f^{-2/3}) \ . \nn
\eea
Thus, to leading order the potential is given by
\bea
V_{\rm{rel.}} = - 3 e^{K} |W|^2, \label{PotentialRelativistic1}
\eea
which is negative for any choice of superpotential. The term arising from eliminating $F$ is sub-leading. It is evident, therefore, that inflation cannot occur since a phase of de-Sitter-like expansion requires a positive energy density in the universe. Thus, supergravitational relativistic DBI inflation with a single chiral superfield does not work!

\section{DBI Inflation from Coupling to a Second Superfield} \label{SectionSingle}

We have shown that, in the relativistic limit, the supergravitational DBI theory of a single chiral supermultiplet $\Phi$ has a negative potential energy and, hence, inflation cannot occur. Let us now extend this theory by coupling it to a second chiral superfield $S$ with component expansion
\be
S = B + \Th^\a \Th_\a F_B.
\ee
Here $B$ is a complex scalar and $F_B$ the complex auxiliary field associated with $S.$ We will assume that this second field has a two-derivative action\footnote{One could equally well assume that it also has higher-derivative kinetic terms, but that they are unimportant in the vacuum. For simplicity, we will not pursue this option here.}. Then, choosing a K\"{a}hler potential such that 
\bea && K_{,AA^*}=1 \, , \label{pen1} \\ && K_{,AB^*} = 0 = K_{,A^*B} \, , \label{pen2} \eea 
and after the same manipulations as in the previous section--for example, Weyl rescaling the action and eliminating the auxiliary fields $b_m,M$--we obtain the Lagrangian
\bea
\frac1e {\cal L} &=& -\frac12 \cR + 3 e^{K} |W|^2 -K_{,BB^*} \pt B \cdot \pt B^* \nn \\ && -\frac{1}{f} \left( \sqrt{1 + 2 f \, \pt A \cdot \pt A^* + f^2 \, (\pt A \cdot \pt A^*)^2 - f^2 \, (\pt A)^2 (\pt A^*)^2} -1 \right)  \nn \\ 
&& + K_{,BB^*} e^{K/3} |F_B|^2 + \frac{}{}e^{2K/3}\big(F_B (D_B W) + F_B^{*} (D_B W)^*\big) \nn \\ &&+  e^{K/3} |F|^2 + \frac{}{}e^{2K/3}\big(F (D_A W) + F^{*} (D_A W)^*\big)  \nn \\ &&  -32 \frac{}{}e^{K/3} |F|^2 \pt A \cdot \pt A^* \, {\cal{T}}  + 16 e^{2K/3} |F|^4 \, {\cal{T}} \ . \label{ActionComponents}
\eea
In this expression, the auxiliary fields $F,F_B$ of the two chiral multiplets have not yet been eliminated. Their equations of motion are given by
\bea
&& F+ e^{K/3} (D_A W)^* + 32 F \, \mac{T} (e^{K/3} |F|^2 - \pt A \cdot \pt A^*) = 0 , \label{EqofMotionFDBI} \\
&& K_{,BB^*} F_B + e^{K/3} (D_B W)^* = 0.
\eea
Note that these equations are not coupled and, thus, $F$ can be eliminated as in the previous section. It is also straightforward to substitute for $F_B,$ since its equation of motion is algebraic and linear. In the non-relativistic limit--that is, when $f$ is small--one obtains the usual potential
\be
V_{\rm{non-rel., 2 \, superfields}} = e^K \big(\, |D_A W|^2 + K^{,BB^*} |D_B W|^2 - 3 |W|^2\big)  .\label{PotentialNonRelativistic}
\ee
However, in the relativistic limit the $|D_A W|^2$ term again is subdominant and the potential becomes
\be
V_{\rm{rel., 2 \, superfields}} = e^K \big( K^{,BB^*} |D_B W|^2 - 3 e^{K} |W|^2 \big)  . \label{PotentialRelativistic}
\ee
Comparing this to expression \eqref{PotentialRelativistic1}, we see that
in the two superfield case a new, positive definite term enters the potential energy!
Hence, by choosing the superpotential appropriately, the overall potential can be made positive along the direction(s) of interest in field space--thus enabling inflation to occur.

We will first be interested in the case where one allows the two real scalars in 
\begin{equation}
A=\frac{1}{\sqrt{2}}(\phi+i\xi)
\label{dood1}
\end{equation}
 to be dynamically relevant. These scalars both have kinetic terms of the DBI form--as is evident, for example, from (\ref{ActionRelativistic1}). Our formalism also implies that, after the potential energy has been chosen for the first scalar, the potential of the second scalar is automatically determined. Moreover, when the K\"{a}hler potential satisfies certain additional requirements--which we derive below--this second scalar can be stabilized. In this case, our construction allows one to obtain an arbitrary positive potential.  Choosing this appropriately leads effectively to a single real component field model of DBI inflation. 

We choose for the superpotential $W$ an Ansatz first used in \cite{Kawasaki:2000yn} and analyzed, in detail, in \cite{Kallosh:2010xz} within the context of ordinary two-derivative supergravity. This Ansatz is
\be
W = S w(\Phi), \label{Ansatz}
\ee
where $w(\Phi)$ is a ``real'' holomorphic function of $\Phi;$ that is, $w(\Phi) = \sum_{\substack{n}} c_n \Phi^n$ with $c_n \in \mathbb{R}.$ The coefficients are chosen to be real for simplicity. The lowest component of $W$ is given by $B w(A)$. On the $B=0$ plane, we have $W=0,D_B W = w(A)$ and, hence, the potential energy \eqref{PotentialRelativistic} becomes
\be
V_{B=0} = e^{K(A,A^*)} K^{,BB^*} |w(A)|^2.
\ee
Here, the K\"{a}hler potential is also evaluated at $B=0$. The $B$ field can always be rescaled so that its kinetic term is canonical (when $B=0$). Correspondingly, we will take $K_{,BB^*} |_{B=0}=1.$ Then the potential further simplifies to
\be
V_{B=0} = e^{K(A,A^*)} |w(A)|^2. \label{Potential2}
\ee
For this expression to be physically relevant, one must ensure that the dynamics is restricted to the $B=0$ plane. That is, the two real scalar fields $b,d,$ defined by
\be
B = \frac{1}{\sqrt{2}} (b + \I d)
\ee 
must be stabilized with zero vacuum expectation values. In an inflationary context, this means that around $b=d=0$ the scalar squared masses $m_b^2,m_d^2$ must be positive and at least as large as the Hubble expansion scale $H^2.$ A straightforward calculation shows that
\bea
m_b^2 &=& \frac{\pt^2 V}{\pt b^2}|_{b=d=0} \nn \\ &=& \Big(\frac{1}{2}\frac{\pt^2 V}{\pt B^2} + \frac{\pt^2 V}{\pt B\pt B^*} +  \frac{1}{2}\frac{\pt^2 V}{\pt B^{*2}}\Big) |_{B=0}\nn \\ &=& -e^{K(A,A^*)} |w(A)|^2 K_{,BBB^*B^*},
\eea
with a similar expression for $m_d^2$. One can assume that, during inflation, the dynamics is dominated by the potential and, thus, the Friedmann equation implies that $V \approx 3H^2.$ Then the requirement that $m_b^2,m_d^2 \gtrsim H^2$ translates into the stability condition
\be
K_{,BBB^*B^*} \lesssim - \frac13. \label{Stability1}
\ee
This condition is analogous to that found in two-derivative supergravity models \cite{Kallosh:2010xz}. It can be satisfied, for example, if the K\"{a}hler potential includes a term $\zeta (BB^*)^2$ with $\zeta \lesssim -1/12.$ 

Now note that for the superpotential (\ref{Ansatz}), $D_A W$ is proportional to $B$ and hence vanishes on the $B=0$ plane. Thus, the potential term $e^K |D_A W|^2$ that becomes subdominant in the relativistic limit, is actually zero on the inflationary trajectory for models of this type. This can also be seen directly from the equation of motion (\ref{EqofMotionFDBI}) for $F$--for the Ansatz (\ref{Ansatz}) the ordinary branch solution for $F$ is simply the trivial solution $F=0$ if we restrict to the $B=0$ plane. In other words, in going from the approximately two-derivative regime to the relativistic DBI regime, the potential does {\it not} change for the models considered here. This special feature is entirely non-trivial, and arises as a direct consequence of the choice (\ref{Ansatz}). It greatly facilitates the analysis of the corresponding inflationary models.
 
Let us now restrict the theory further, so that only a single real scalar field in \eqref{dood1} remains dynamical. For this purpose, choose the K\"{a}hler potential to depend on $\Phi,\Phid$ via the combination $-\frac12 (\Phi - \Phid)^2$ only.
Then, the K\"{a}hler potential will not depend on $\phi.$ Correspondingly, if $\xi$ is now stabilized around $\xi=0$ with a sufficiently high mass, then the dynamics will take place entirely in the $\phi$ direction with the potential 
\be
V_{\phi} = w\left( \frac{\phi}{\sqrt{2}} \right)^2 \ .
\ee 
Thus, any smooth positive potential can be engineered in this way, simply by identifying $w$ with the square root of the desired potential and analytically continuing $w$ to the complex plane \cite{Kallosh:2010xz}. However, for consistency, one must check under what conditions $\xi$ is stabilized. Its mass along the putative inflationary trajectory is given by
\bea
m_\xi^2 &=& \frac{\pt^2 V}{\pt \xi^2}|_{\xi=b=d=0} \nn \\ &=& \Big(-\frac{1}{2}\frac{\pt^2 V}{\pt A^2} + \frac{\pt^2 V}{\pt A\pt A^*} - \frac{1}{2}\frac{\pt^2 V}{\pt A^{*2}}\Big) |_{\xi=B=0}\nn \\ &=& -w w'' + w'{}^2 + 2 w^2(1- K_{,AA^*BB^*}), \label{StabilityCalculation}
\eea
where $w' = w_{,A}|_{\xi=0}.$ This mass is identical to that obtained in two-derivative supergravity theories \cite{Kallosh:2010xz}. A working model of single real component field DBI inflation must then satisfy $m_\xi^2 \gtrsim H^2$--otherwise perturbations in the $\xi$ field also become relevant. When $w''/w$ and $(w'/w)^2$ are small (bearing in mind  that for DBI inflation they need not be as small as for two-derivative inflation), this translates into the requirement
\be
K_{,AA^*BB^*} \lesssim \frac56. \label{Stability2}
\ee
An example of a K\"{a}hler potential satisfying all of the above assumptions and stability constraints was discussed in \cite{Kallosh:2010xz}. Here, we will simply repeat it for specificity. It is given by
\be
K = -\frac12 (\Phi - \Phid)^2 + SS^\dag + \zeta(SS^\dag)^2 + \frac{\gamma}{2}SS^\dag (\Phi - \Phid)^2 \label{KExample}
\ee
with $\zeta \lesssim - 1/12$ and $\gamma \gtrsim 5/6.$

\section{Coupling to Additional Chiral Superfields} \label{SectionMulti}

DBI inflation was inspired by string theory, and is of importance because it has a more direct link to microphysics than most inflationary models. The higher-derivative terms play a crucial role in DBI theories, since they lead to the speed limit (\ref{SpeedLimit}). They also imply the generation of significant equilateral non-Gaussianity \cite{Silverstein:2003hf,Alishahiha:2004eh}. Interestingly, models of single real scalar field DBI inflation are already tightly constrained by current observations--precisely because of the constraints imposed by the underlying microphysics. Such models could be ruled out in the near future \cite{Baumann:2006cd,Lidsey:2007gq,Kobayashi:2007hm,Bean:2007eh}. However, restricting to a single real scalar field is not necessary within a string theory context. For example, many DBI models that have been considered focus on a D3-brane moving along a warped throat of an internal Calabi-Yau manifold. The radial direction is typically identified with the inflaton. By construction, however, such models naturally have multiple real scalar fields, with the angular directions in the Calabi-Yau space providing the additional degrees of freedom \cite{Langlois:2008qf}. Hence, it is of interest to also study multi-field models of DBI inflation. For such theories, the constraints arising from the comparison with observational data are typically less severe. An interesting recent example is provided in  \cite{Kidani:2012jp}, which is in agreement with all current observations, but where significant non-Gaussianities of both local and equilateral type are predicted.

The models studied in the previous section, if the second real scalar $\xi$ is {\it not} stabilized, can be regarded as two real scalar field models. This can be achieved by removing restriction \eqref{Stability2} on the K\"ahler potential. However, the form of the potential (\ref{Potential2}) is then rather restrictive.
We found that an essentially arbitrary positive potential could be obtained in the purely $\phi$ direction by choosing $w(A)$ appropriately. But, given $w(A),$ the potential for the second field $\xi$ is then determined at the same time. Hence, there is a risk that the second direction spoils the suitability of the potential for inflationary dynamics \cite{Ovrut:1983my}. It turns out that more flexibility in constructing multi-real-scalar-field potentials can be obtained by coupling our theory to a third chiral superfield $\Psi$, with component expansion
\be
\Psi = C + \Th^\a \Th_\a F_C.
\ee
We will assume that $\Psi,$ just like $S,$ does not appear with higher-derivative kinetic terms in the Lagrangian. Then, in analogy with $F_B$ above, the auxiliary field $F_C$ is easily eliminated. Furthermore, in addition to conditions \eqref{pen1},\eqref{pen2} we restrict the K\"ahler potential to satisfy
\begin{equation}
K_{,BC^{*}}=0=K_{,B^{*}C} \ .
\label{pen3}
\end{equation}
In the relativistic limit, the potential now becomes
\be
V_{\rm{rel.,3 \, superfields}} = e^K \big(K^{,BB^*} |D_B W|^2 + K^{,CC^*} |D_C W|^2 - 3 |W|^2\big) \ .
\ee
In the non-relativistic limit there would be an additional term $e^K |D_A W|^2.$ 

When considering multiple fields, inflationary models of the so-called {\it hybrid} type are of particular interest. In such theories, inflation occurs along a direction that becomes unstable at a certain field value. At this point, the inflationary trajectory makes a turn in field space, following the locally steepest direction to a true minimum of the potential. To obtain such models in two-derivative supergravity, a superpotential of the form
\be
W = \Phi (a_1 \Psi^2 - a_2)
\ee
has been used \cite{Copeland:1994vg,Lyth:1998xn}, where $a_{1,2}$ are constants\footnote{In the context of supersymmetric GUT models, this can be generalized to a pair of conjugate chiral fields $\Psi,{\Psi}^c,$ which transform non-trivially under the action of a gauge group--see, for example \cite{Linde:1997sj,Yamaguchi:2011kg}. In this case, one may choose a superpotential of the form $W = \Phi (a_1 \Psi {\Psi}^c - a_2).$}. In the present context, this approach does not work! The reason is that the $D_A W$ terms, which are needed to obtain the desired potential, are subdominant in the relativistic regime. However, instead of coupling $\Psi$ directly to $\Phi,$ one can couple $\Psi$ to $S$ instead--while continuing to couple $S$ to $\Phi$ as in the previous section. Thus, we consider the superpotential
\be
W = S  w(\Phi,\Psi), \label{Ansatz2}
\ee
where $w$ is now a holomorphic function of $\Phi$ and $\Psi,$ and where we will assume that in the double Taylor series expansion of $w$ in $\Phi,\Psi$ only real coefficients occur. The lowest component of this superpotential is $B w(A,C).$ Note that $D_A W,\,$ $D_C W$ and $W$ itself are all proportional to $B$. Hence, if the stability condition (\ref{Stability1}) holds, then the dynamics once again takes place entirely on the hypersurface $B=0.$ The potential is then generated solely by the $D_B W$ term, and reduces to
\be
V_{B=0} = e^{K(A,A^*,C,C^*)} |w(A,C)|^2.
\ee
Note that since $D_A W$ is zero in the field space region of interest, the corresponding ordinary branch solution for $F$ is once again simply $F=0$. Therefore, the potential is {\it always} given by the above expression, whether the higher-derivative DBI terms are important or not.

\begin{figure}[h!]
\begin{center}
\includegraphics[width=0.7\textwidth]{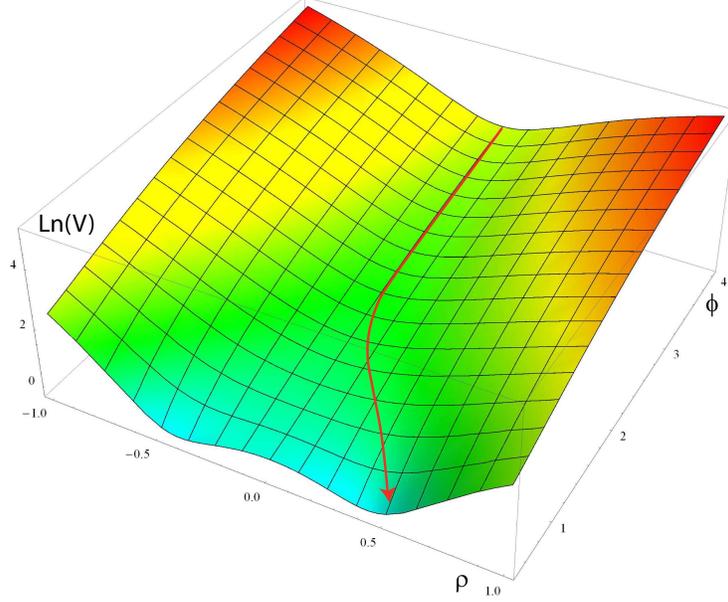}
\caption{\label{Figure1} {\small This graph depicts the field space trajectory in a potential of the hybrid form (\ref{Potential4}), with $a_0=a_2=1, a_1 = 10, a_3=4$. For illustration purposes, we have plotted the logarithm of the potential rather than the potential itself. The trajectory first evolves along the $\phi$ direction with a $\phi^2$ potential, and then turns as the transverse direction becomes unstable. Inflation ends as the trajectory reaches a true minimum of the potential at $\phi=0,\rho=1/2$ (or $-1/2$). For the models we have constructed, the kinetic terms are also of  a ``hybrid'' type: the $\phi$ field evolves according to a DBI kinetic term, while $\rho$ is governed by a standard two-derivative kinetic term.}}
\end{center}
\end{figure}

Similar to the analysis of Section \ref{SectionSingle}, we now investigate whether one can further restrict the dynamics to the two directions $\phi = \sqrt{2} \Re (A)$ and $\rho=\sqrt{2} \Re (C).$ For this to be possible, we must ensure that the directions $\xi = \sqrt{2}\Im(A)$ and $\tau=\sqrt{2}\Im(C)$ are stabilized when $B=0$. Assuming that the K\"{a}hler potential depends only on the combinations $-\frac12(\Phi - \Phid)^2$ and $-\frac12(\Psi - \Psi^\dag)^2,$ an analogous calculation to (\ref{StabilityCalculation}) shows that the corresponding masses are given by
\bea
m_\xi^2 &=& -w w_{,AA} + w_{,A}{}^2 + 2 w^2(1- K_{,AA^*BB^*}),\\
m_\tau^2 &=& -w w_{,CC} + w_{,C}{}^2 + 2 w^2(1- K_{,CC^*BB^*}),
\eea 
where all terms are evaluated at $B=0.$ Dynamical stability during inflation is guaranteed if these masses are larger than the Hubble scale. As above, neglecting $w_{,AA}/w,w_{,CC}/w,w_{,A}^2/w^2$ and $w_{,C}^2/w^2$, we obtain the following requirements on the K\"{a}hler potential;
\bea
K_{,AA^*BB^*} &\lesssim& \frac56, \\
K_{,CC^*BB^*} &\lesssim& \frac56.
\eea
Under these conditions, the potential energy further simplifies to
\be
V_{\phi,\rho} = \left(w(\frac{\phi}{\sqrt{2}},\frac{\rho}{\sqrt{2}})\right)^2. \label{Potential3}
\ee
An example of a K\"{a}hler potential satisfying all of the assumptions and constraints above is given by an extension of (\ref{KExample}),
\bea
K &=& -\frac12 (\Phi - \Phid)^2 + SS^\dag + \zeta(SS^\dag)^2 + \frac{\gamma_1}{2}SS^\dag (\Phi - \Phid)^2 \nn \\ && -\frac12 (\Psi-\Psi^\dag)^2 + \frac{\gamma_2}{2}SS^\dag (\Psi-\Psi^\dag)^2,
\eea
with $\zeta \lesssim - 1/12,\gamma_1 \gtrsim 5/6, \gamma_2 \gtrsim 5/6.$ In this case, four out of the six real scalars fields are stabilized. The two remaining scalars are dynamical fields, moving in an essentially arbitrary potential given by (\ref{Potential3}). For example, a typical hybrid potential can be obtained by choosing 
\be
W_{\rm{hybrid}} = S \sqrt{a_0^2 \big( 2\Phi^2 +4 a_1 \Phi^2 \Psi^2 + (a_2- 2 a_3 \Psi^2)^2\big)},
\ee 
with real positive constants $a_{0,1,2,3}$. This  leads to the potential energy
\be
V_{\rm{hybrid}} = a_0^2 \big(\phi^2 + a_1 \phi^2 \rho^2+ (a_2-a_3 \rho^2)^2\big). \label{Potential4}
\ee
For $\phi > \sqrt{2a_2 a_3/a_1}$, inflation takes place along the $\rho=0$ line with potential $a_0^2 (\phi^2+a_2^2).$ For $\phi < \sqrt{2a_2 a_3/a_1},$ the transverse direction turns over, and two new minima now arise at $\phi=0,\rho = \pm \sqrt{a_2/a_3},$ at which points the potential vanishes--see Fig. \ref{Figure1}. This example illustrates how two-field potentials can be engineered by choosing the superpotential appropriately. One special feature of the models considered here is that the kinetic terms are also ``hybrid''--in the sense that $\phi$ has a higher-derivative DBI action, while the kinetic term for $\rho$ is a canonical two-derivative one.

We should add that for models where the additional fields transform non-trivially under a gauge group, radiative corrections must be taken into account \cite{Yamaguchi:2011kg}. A more thorough analysis is then required on a case by case basis. Additionally, we would like to note that in all of our constructions, we have looked only at the inflationary sector of the theory. In a more complete setting, it is important to check that the interactions with other sectors do not spoil the inflationary dynamics \cite{Hardeman:2010fh}. Of course, this issue must also be analyzed with a specific model at hand.

\section{Conclusions} \label{SectionConclusions}

One of the most important problems in cosmology is to find a scenario for the early universe that is not only in agreement with observations, but is also rooted in a sensible microphysical theory. Only in this way can cosmology and particle physics be united, and a consistent theory of our universe be obtained. While still far from this goal, we have analyzed a small aspect of the problem in this paper--showing how to construct models of DBI inflation in four-dimensional ${\cal{N}}=1$ supergravity. 

Our recent supergravity analysis of higher-derivative actions showed that if one tries to construct a model of DBI inflation from a  single chiral superfield, it is bound to fail--since the potential becomes negative when the higher-derivative terms become important. In this paper, we circumvented this problem by coupling the theory to one or more additional chiral superfields. In fact, the construction in Section \ref{SectionMulti} can be generalized to an arbitrary number $N$ of chiral superfields--each with two-derivative kinetic terms and appropriately constrained K\"ahler potential--and considering a superpotential of the form
\be
W = S w(\Phi^1,\Phi^2,\dots,\Phi^N) .
\ee 
Then, not only can the potential energy be positive but one can construct a wide range of potential functions for the original DBI scalar $\phi = \sqrt{2} \Re(\Phi^1)$ and $N-1$ additional real scalars. The remaining real scalars, that is, the two making up the lowest component of $S$ and one scalar in the lowest component of all the other chiral superfields, can be stabilized with masses above the Hubble scale if the K\"{a}hler potential satisfies certain requirements discussed in the text. 
Our analysis can be viewed as a ``proof-in-principle'' that models of multi-real-scalar-field DBI inflation can be constructed in ${\cal{N}}=1$ supergravity.

A crucial feature of the analysis of chiral superfields with higher-derivative actions is that, via the elimination of the auxiliary fields, the potential energy  generically depends not only on the superpotential, but on the strength of the higher-derivative terms as well. Thus, in general, the potential changes during the dynamical evolution. In this paper, we have shown that, for the constrained K\"ahler potentials and superpotentials above, this turns out not to be the case. The contributions to the potential that depend on the higher-derivative terms vanish in the region of field space of dynamical interest. Thus, the potential remains unchanged as the higher-derivative terms become large or small. This feature considerably simplifies the study of the models considered here, and renders them more accessible for deriving their predictions for cosmological observations. We hope to pursue this topic in the near future.

Our construction illustrates that it is far from straightforward to realize DBI inflation in ${\cal{N}}=1$ supergravity. We have shown one way in which 
the desired positive potentials can be obtained from an effective model-building point of view. It is interesting to ask whether there exist other ways of realizing DBI inflation within the context of supergravity. More importantly, however, is the question of whether or not such constructions can be obtained from a full-fledged string compactification, or from some other fundamental theory of particle physics. These are pertinent questions for future research.

\section*{Acknowledgments}

M.K. and J.L.L. gratefully acknowledge the support of the European Research Council via the Starting Grant numbered 256994. B.A.O. is supported in part by the DOE under contract No. DE-AC02-76-ER-03071 and the NSF under grant No. 1001296.

\bibliographystyle{unsrtabbrv}

%********************************
  
  %\cite{Gray:2007zza}
%\bibitem{Gray:2007zza} 
%  J.~Gray, A.~Lukas and B.~Ovrut,
  %``Perturbative anti-brane potentials in heterotic M-theory,''
 % Phys.\ Rev.\ D {\bf 76}, 066007 (2007)
%  [hep-th/0701025].
  %%CITATION = HEP-TH/0701025;%%
  
  %\cite{Buchbinder:2002pr}
%\bibitem{Buchbinder:2002pr} 
%  E.~I.~Buchbinder, R.~Donagi and B.~A.~Ovrut,
  %``Vector bundle moduli superpotentials in heterotic superstrings and M theory,''
%  JHEP {\bf 0207}, 066 (2002)
%  [hep-th/0206203].
  %%CITATION = HEP-TH/0206203;%%
  
  %\cite{Lima:2001nh}
%\bibitem{Lima:2001nh} 
 % E.~Lima, B.~A.~Ovrut and J.~Park,
  %``Five-brane superpotentials in heterotic M theory,''
 % Nucl.\ Phys.\ B {\bf 626}, 113 (2002)
 % [hep-th/0102046].
  %%CITATION = HEP-TH/0102046;%%

%**********************************  
\bibliography{master}
\end{document}